# Local Turbulence: Effects and causes

Olivier Lai[*,a], Kanoa Withington[b], Romain Laugier[a], Mark Chun[c]

[a]Université Côte d'Azur, Observatoire de la Côte d'Azur, CNRS, Laboratoire Lagrange, Bd de l'Observatoire, CS 34229, 06304 Nice cedex 4, France; [b]Canada France Hawaii Telescope, 65-1238 Mamalahoa Hwy, Kamuela HI 96743, [c]Institute for Astronomy, University of Hawaii, 640 N. A'ohoku Place, Hilo, HI 96720

## ABSTRACT

Dome seeing is a known source of image quality degradation, but despite tremendous progress in wavefront control with the development of adaptive optics and environmental control through implementation of dome venting, surprisingly little is known about it quantitatively. We have found evidence of non-Kolmogorov dome turbulence from our observations with the imaka wide field adaptive optics system; PSFs seem to indicate an excess of high spatial frequencies and turbulence profiles reveal turbulence at negative conjugations. This has motivated the development of a new type of optical turbulence sensor called AIR-FLOW, Airborne Interferometric Recombiner: Fluctuations of Light at Optical Wavelengths. It is a non-redundant mask imaging interferometer that samples the optical turbulence passing through a measurement cell and it measures the two-dimensional optical Phase Structure Function. This is a useful tool to characterise different types of turbulence (e.g. Kolmogorov, diffusive turbulence, etc.). By fitting different models, we can determine parameters such as $C_n^2$, $r_0$, $L_0$ or deviation from fully developed turbulence. The instrument was tested at the Canada France Hawaii Telescope, at the University of Hawaii 2.2-meter telescope (UH88") and at the Observatoire de la Côte d'Azur. It is ruggedised and sensitive enough to detect changes with different dome vent configurations, as well as slow local variations of the index of refraction in the UH88" telescope tube. The instrument is portable enough that it can be used to locate sources of turbulence inside and around domes, but it can also be used in an operational setting without affecting observations to characterise the local optical turbulence responsible for dome seeing. Thus, it could be used in real-time observatory control systems to configure vents and air handlers to effectively reduce dome seeing. We believe it could also be a tool for site surveys to evaluate dome seeing mitigation strategies in situ.

## 1. INTRODUCTION

Ground Layer adaptive optics[1] aims at providing adaptive correction of atmospheric turbulence over fields as wide as possible. It does so by correcting turbulence at or close to the pupil, which is common to the entire field. In its most basic form, the principle is the following: multiple wavefront sensors pointed at guide stars across the field are averaged to isolate the ground layer turbulence, and the correction is applied to a deformable mirror conjugate to the pupil. However, in its practical implementation, the limited number of guide stars implies that the uncorrelated turbulence of the free atmosphere is not perfectly averaged: the variance of this undesirable signal decreases as *1/N* where *N* is the number of guide stars, and consequently degrades the correction across the entire field as it is fed back into the correction via the deformable mirror. Also in practice, it is not possible perfectly conjugate a deformable mirror to the pupil: it must either be tilted (so that the beam can avoid the collimator after reflection on the deformable mirror), or in the case of an adaptive secondary mirror (ASM), there will be a small, but not inconsequential misconjugation which will limit the off-axis performance[2]. We studied these effects extensively through simulations, using a Monte Carlo code developed specifically to incorporate wide field GLAO effects (`instant_GLAO`), but found that the simulations results were almost completely dependent on the input perturbation (turbulence strength and profile.) Thus, to study these effects beyond simulations, we developed the imaka GLAO prototype on the UH88" telescope, as a testbench and demonstrator[3,4]. It consists of an optical relay which transmits a 0.4°x0.3° field of view for (up to) fixed 5 wavefront sensors, a location for a deformable mirror at a pupil plane (although tilted with respect to it), and a diffraction limited science field of 10'x10'. The goal of this GLAO system is to study quantitative gains on photometry and astrometry as such gains depend much more critically on the Noise

---

[*]olivier.lai@oca.eu

Equivalent Area (NEA) than, for example, more traditional metrics used to qualify GLAO performance such as FWHM. The NEA is much more sensitive to the morphology of the PSF[5], and the relative strength of the wings versus the peakiness of the core. As such it is very sensitive to real world effects such the dome seeing, or the outer scale of the Ground Layer (GL) which are difficult to include in simulations, for lack of input data.

**1.1 On-sky image improvements with imaka**

Imaka saw its first light in October 2016, using a commercially available Finger Lakes Instrumentation (FLI) CCD camera with 8000x6000 pixels. Focal plane data acquired during subsequent runs was comprehensively analyzed and published[6]. Here we show only a few results to highlight some aspects of PSF morphology and how they suggest the presence of local turbulence. We first show a comparison of FWHM and NEA for data obtained on the night of January 12$^{th}$, 2017 (Figure1). The open loop, uncorrected empirically measured FWHM is on the order of 0.8" that night, and when we apply the GLAO correction, this drops to approximately 0.4", a gain of a factor 2. The NEA can be thought of as the area of the optimal aperture for signal-to-noise in the background limited case. If the PSF had a cylindrical morphology, we would expect NEA = $\pi$ (FWHM/2)$^2$. If the morphology of the PSF changes, this relation may not hold true anymore; for example, if more energy is spread out into the wings, the NEA will increase, while the FWHM may remain more or less constant. In the case of imaka we find that the gain in NEA and the gain in FWHM are both a factor 2, so that NEA $\propto$ FWHM.

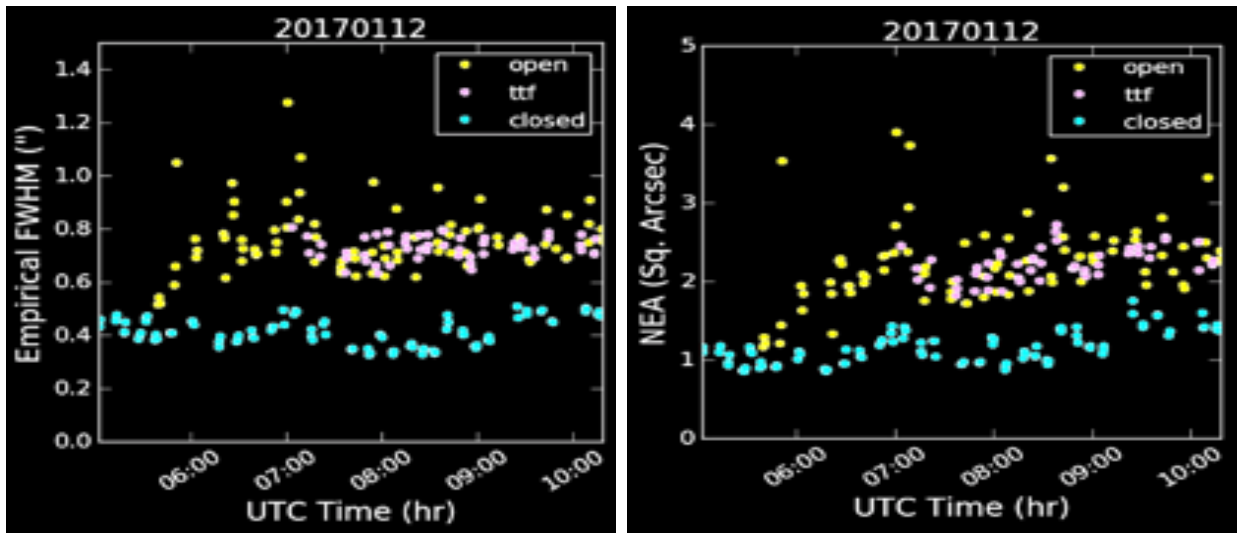

Figure 1. Left: the measured FWHM of the PSF averaged across the field in open loop, tip-tilt loop and closed loop for the night of January 12$^{th}$, 2017. Right: The Noise Equivalent Area for the same PSFs. Note that the gain in NEA (which is an area) is not proportional to the gain in FWHM$^2$.

Light scattered far from the core of the PSF is due to high spatial frequencies, small scale defects in amplitude or phase in the pupil. It is also well known that introducing an outer scale to Kolmogorov turbulence (von Kalman turbulence) decreases the relative contribution of low spatial frequencies, and turbulence inside a telescope dome cannot have an outer scale larger than the dome itself. Obstacles in the airflow, due to the topography (or the dome and telescope structures themselves) will have a finite outer scale of a size comparable to that of the obstacle. Finally, diffusive turbulence (when a denser fluid is above a less dense one, leading to Rayleigh-Taylor instabilities in the mixing region) is also predominantly a high spatial frequency phenomenon (random walk turbulence). We therefore hypothesize that the power spectrum of the turbulence that we are trying to correct does not follow the Kolmogorov hypothesis and instead carries an excess of high spatial frequencies due to local turbulence.

**1.2 PSF Moffat fitting**

To develop the PSF morphological analysis further, we have fitted them by a Moffat function. This allows the PSFs to be parametrized with only two parameters: the FWHM and the ß exponent. The latter describes how peaked the profile is: when ß $\rightarrow \infty$, the profile is Gaussian, while ß=1 is a Lorentzian profile, with much stronger wings (in fact a diffraction limited Airy pattern has an envelope that decreases as *1/r$^2$* for which ß=1). We find in the literature[7] that observed PSFs at various telescopes have a value of ß between 4 and 5. When an image is predominantly degraded by vibrations, the profile becomes closer to a Gaussian and ß increases correspondingly.

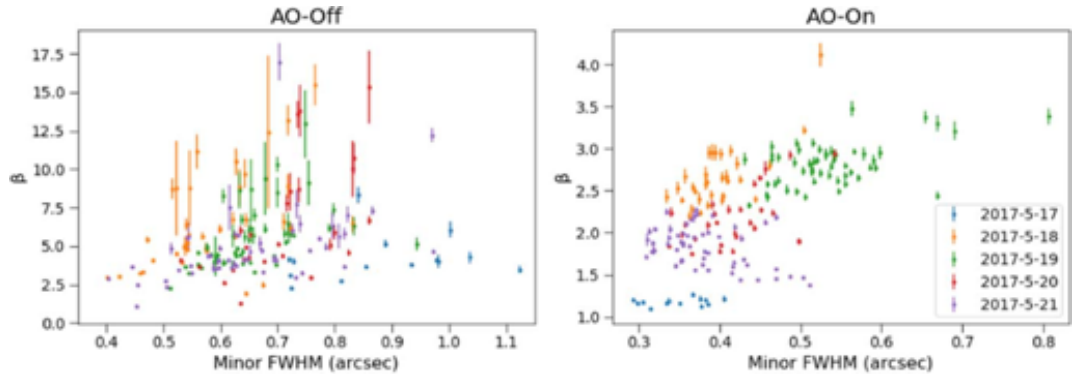

Figure 2.: Moffat parameters for May 2017 observing run, open loop (left) and closed loop (right). Median values of ß in open loop are around 4.8 but some values are much larger when vibrations are present. In closed loop, the median is closer to 2, but is as low as 1.1 during the night of the 17$^{th}$.

Data obtained during Our May 2017 run shows that the median ß obtained in open loop is around 4.8 (with some values much higher due to telescope vibrations), as shown in Figure 2. In closed loop values are systematically lower, with a median closer to 2, but some value as low as 1.1, implying much stronger wings in the corrected PSF. We have computed synthetic PSFs for various wavelengths for Kolmogorov and von Kalman phase structure function with varying $r_0$ (and $L_0$) and have fit them with Moffat profiles, as show on Figure 3 (Left). We find a small dependence of ß with $r_0$ (ß decreases as $r_0$ increases), and a strong dependence of ß with $L_0$ (ß decreases when $L_0$ decreases). This implies that the closed loop PSF is morphologically equivalent to an open loop PSF with a much smaller outer scale. This is confirmed if we directly fit our observed PSFs with our synthetic von Kalman PSF, letting $r_0$ and $L_0$ vary as free parameters in the fitting process[8], as shown on Figure 3 (Right). This raises the question as to whether this morphology is characteristic to the GLAO PSF, and whether this is due to an excess of high spatial frequencies in the Ground Layer, that a low order AO system is unable to correct.

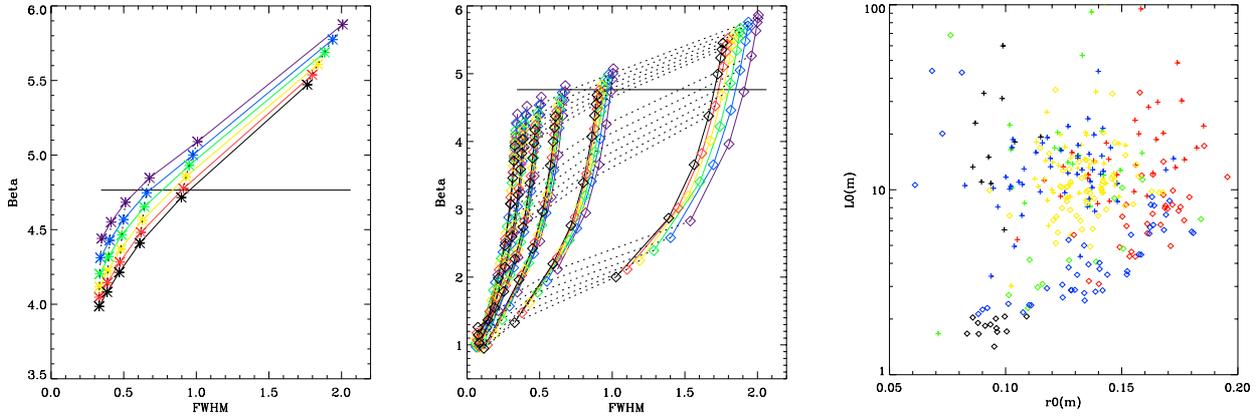

Figure 3: Left, Moffat parameters FWHM and ß for synthetic Kolmogorov PSFs showing a decrease of ß as $r_0$ increases. Different colors indicate different wavelengths ranging from 500nm to 1μm. Middle, including an outer scale for the same values of $r_0$ produces a large decrease of ß. Right, r0, L0 parametric fitting of our observed PSF shows that closed loop (diamonds) have roughly similar $r_0$ than open loop (crosses), but systematically lower $L_0$. In this plot, different colors indicate different nights.

## 1.3 Tube seeing

When observing with imaka, we also collect wavefront sensor and deformable mirror telemetry. We can use this to compute the turbulence profile at the time of observation, using different wavefront slope covariance matrices in a method similar to SloDaR. Nothing prevents us from computing wavefront slope correlations at negative altitudes, which corresponds to turbulence inside the telescope tube[8]. Furthermore, tube turbulence should have a characteristic signature due to the triple pass of the beam through turbulent layers inside the tube in a Cassegrain system: A first time as the beam reaches the primary mirror, at a small but positive conjugation, a second time as the beam converges towards the secondary mirror at a small but negative conjugation, and a third time on its way from the secondary to focus at a larger negative conjugation. As the beam becomes smaller at each pass, the $D/r_0$ is smaller, so the strength of each echo should also decrease.

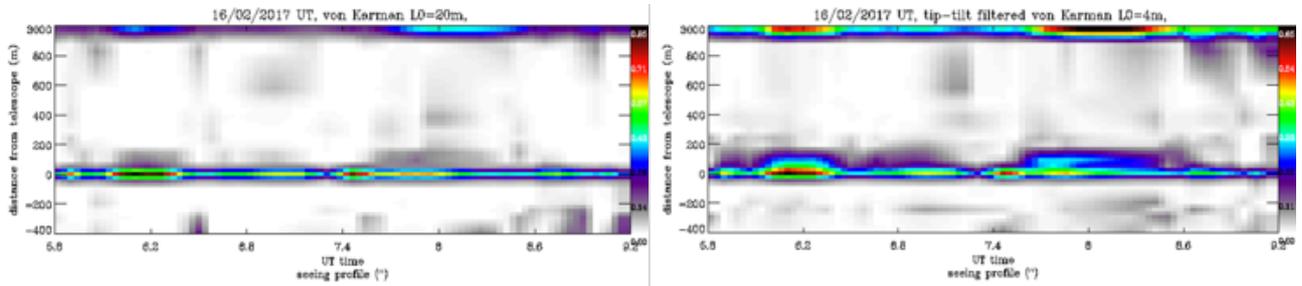

Figure 4: Vertical turbulence profile for February 16$^{th}$, 2017 for the atmosphere and inside the telescope tube (negative conjugations). Left shows the full data including tip-tilt, which is more sensitive to vibrations and guiding errors, while the right panel shows the same algorithm on tip-tilt filtered data, which is more reliable, but less sensitive to large outer scale turbulence. In the right panel, structure, layering and possible echoes are visible in the latter half of the night.

For the UH88" telescope, the secondary mirror is conjugate to -19m and the hole in the primary is conjugate to -400m (a layer 25cm above the primary is conjugate at -300m). Figure 4 shows turbulence profiles for the night of February 16$^{th}$, 2017. The vertical resolution is not sufficient to resolve the first two echoes, but we can see turbulence at negative conjugation, inside the tube, in the second half of the night. The event at 7.4UT in particular shows a strong peak of turbulence at the ground with an echo at -400m, which could be indicative of turbulence very close to the primary mirror.

All of the above are reasons to believe that tube and dome seeing play an important part in the image degradation of any telescope, but especially in the case of GLAO, where the gain in quantitative performance may not be as substantial as simulations lead to believe. In fact, there is a discrepancy of up to 0.4" in the telescope's focal plane IQ when compared to the MKAM (Mauna Kea Atmospheric Monitor) DIMM, which is most likely attributable to local effects in and around the dome. Local turbulence effects have also been known to be detrimental in high dynamic range systems with the low wind effect, seen on the VLT with SPHERE and on Subaru with SCExAO, whereby radiative supercooling of the secondary mirror spiders is thought to produce cascades of cold air that introduce discontinuities and a gradient across the pupil's quadrants. This raises questions as to the environment of the telescope with respect to self-generated turbulence: for example, telescope trusses and top rings could also supercool and with a little bit of wind inside the dome, introduce optical turbulence into the telescope beam. The power spectrum of such turbulence is very different from Kolmogorov, but most (if not all) simulations of adaptive optics, which are often used at the design level, do not take this into account and assume Kolmogorov or von Kalman turbulence. This raises the question of how useful or trustworthy such simulations really are! This is obviously also true for the temporal characteristics of the turbulence and its correction.

The problem of dome turbulence is therefore very complex, and ever-evolving trends in dome designs illustrate that this problem is far from being solved. Hydrodynamical simulations are sometimes used to assist with the design of domes that minimize self-induced turbulence, but it should be noted that mapping out mechanical turbulence may not be a sufficient criterion, since temperature differences are also needed to imprint optical turbulence onto mechanical turbulence, and dome environments are highly complex from a thermal point of view. Besides inside/outside temperature differences and mixing, heat sources from electronic components and hydrostatic bearings, radiative cooling of structures, warm air trapped in telescope tubes or primary mirror cells (or seeping through telescope segment gaps) all need to be taken into account. For all these reasons, we wanted to develop a heuristic and empirical approach to make progress in the measurement, and eventual control of self-generated turbulence inside and around telescope domes.

## 2. AIRFLOW

Ideally, we would have developed a "turbulence camera", similar to Schlieren photography but able to work on extended scenes without needing large optics, and work is ongoing to develop concepts that use other forms of coherence to measure the deviation of light beams introduced by turbulence across the region of interest. But as an interim solution, we decided to develop a small, portable and reliable localized optical turbulence sensor with which we would be able to map out optical turbulence in a volume by scanning it. The sensor was named AIR-FLOW[9] (Airborne Interferometric recombiner: Fluctuations of Light at Optical Wavelengths) and works on the principle of non-redundant masking, allowing to measure the variance of optical path differences introduced by turbulence at different separations and thus to measure the optical Phase Structure Function, which is a very useful tool to characterize turbulence: The Phase Structure Function for Kolmogorov turbulence is proportional to $r^{5/3}$ (and any flattening at larger separations can be attributed to a finite outer scale), diffusive turbulence is proportional to $r$, and tip-tilt (produced when two fluids with different indices of refraction slosh around without mixing) is proportional to $r^2$. The principle and implementation of a prototype sensor are shown in Figure 5.

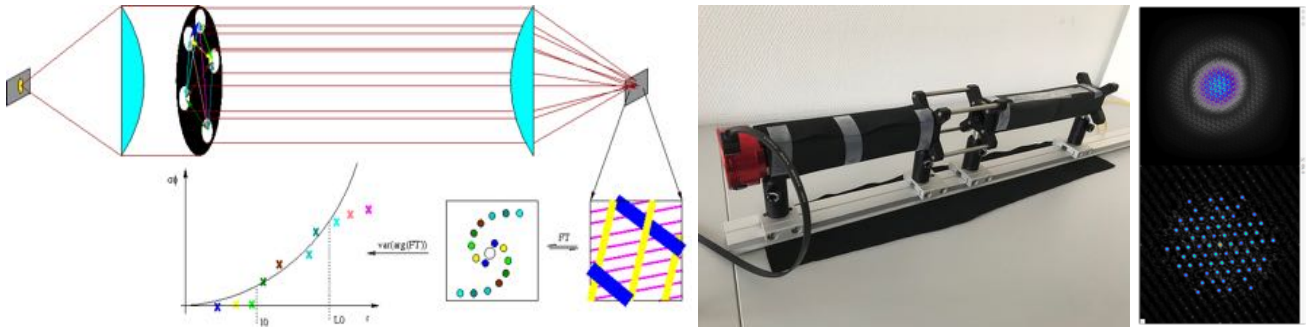

Figure 5, Left, principle of the optical turbulence sensor: a source is reimaged onto a detector after the beam has propagated through a turbulence sensing cell, where the beam is parallel, and a non-redundant mask selects only a few beams with unique baseline vectors. The image recorded on the detector is the complex interferometric PSF (show on the top right), the Fourier transform of which (shown bottom right) allows to extract the phase variance for each baseline. A photo of the device is shown in the middle panel.

The camera was chosen for its small pixel size (ZWO ASi178mm with 2.5µm pixels), and the light source is a single mode fiber fed by a laser diode. All components are commercial off-the-shelf apart from the non-redundant mask that was 3D printed. Data acquisition can be done using any computer with USB3 capability and we are developing an onboard computer based on an ODROID-XU3 single board computer. A data sequence consists of a few thousand sub-rastered images obtained at relatively high frame rate (up to 200Hz) and a data analysis tool with graphical user interface was developed and is shown on Figure 6.

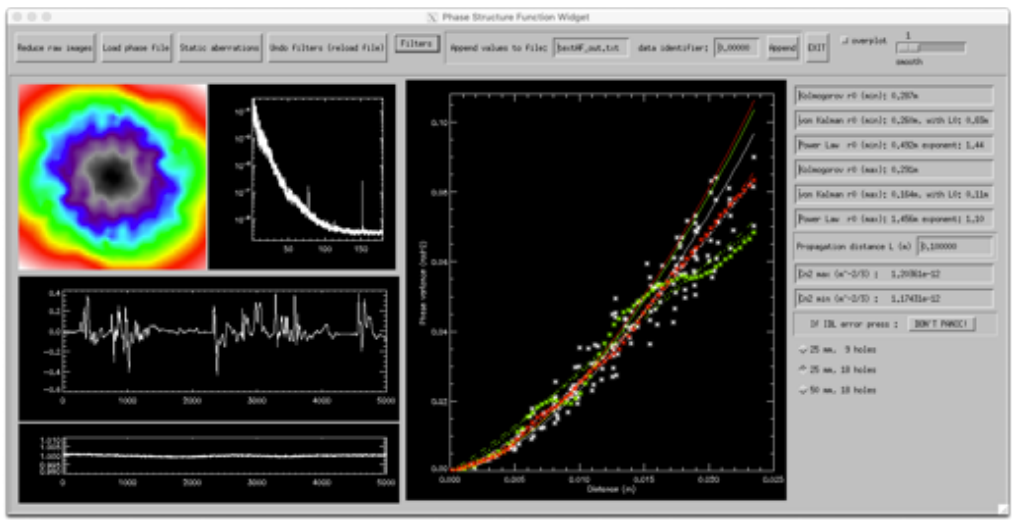

Figure 6, data processing and analysis tool. The left panel shows the 2D phase structure function (top left), the temporal power spectrum of the average phase (top right), the average phase (middle) and the relative flux (bottom). The right panel shows the phase structure function with different fits and values associated with them.

The instrument is compact, reliable and ruggedized for field use and provides direct measurements of optical turbulence. Because the latter is a stochastic process with inherent variations, we decided to not spend too much energy into improving it, although there are obviously many ways that it could be. Currently the beam diameter is 25mm, which makes it relatively insensitive to the outer scale. However, the pixel sampling on the camera requires the use of a collimator with an f-ratio of 8. Making the beam diameter larger would make the length of the instrument much more cumbersome and sensitive to vibrations and flexures. We are however investigating the possible use of cameras with 1.1µm pixels which would allow the use of larger beams while remaining fairly compact.

## 3. FIELD TESTS

After extensive laboratory testing, we decided to try out the sensor in an operational environment, since the amount of localized turbulence inside domes was not known and would determine whether we had the required level of sensitivity.

### 3.1 CFHT Daytime testing

We installed the sensor on one of the serrurier trusses of the CFH Telescope on October 4th, 2018, as shown on Figure 7. We chose this location due to its proximity to the telescope beam and because it was easily accessible with a scissor lift.

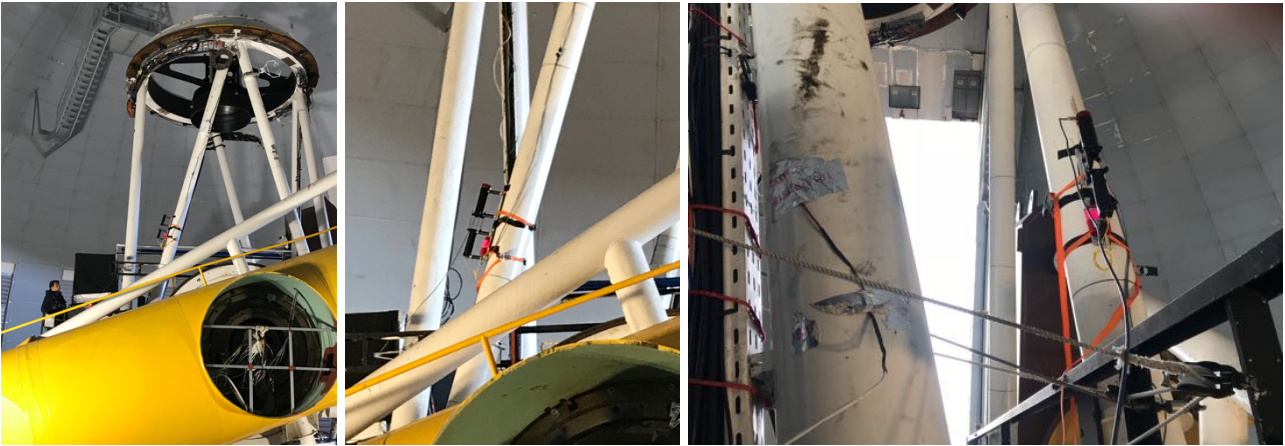

Figure7. AIRFLOW mounted on the CFHT serrurier truss, close to the telescope beam.

We took measurements during the day as we wanted to operate the slit and vents without affecting observing, although daytime turbulence conditions can be expected to be different than at night, especially as the ground was being vigorously heated, and a ~20mph wind was blowing from the North-East. Nonetheless, we went through the following sequence of events: dome closed, slit open, East (upwind) vents open, East & West vents open and East vents closed, West (downwind) vents open and recorded $C_n^2$ values for each, shown on Figure 8. Baseline measurements with the dome closed produce values of $C_n^2$ of a few $10^{-16}m^{-2/3}$ to $10^{-15}m^{-2/3}$. Opening the slit (which was pointed North, perpendicular to the wind direction) did not significantly change the $C_n^2$ apart from some transient buffeting that could also be felt, anecdotally when standing nearby. When the East vents were opened, the $C_n^2$ increased significantly (to $10^{-13}m^{-2/3}$), but seemed to decrease over time, as if the dome was starting to thermalize, with still a few points remaining high, possibly as outside turbulence entered the dome. When the West vents were opened, the flow of air was unimpeded and values of $10^{-13}m^{-2/3}$ were dominant; it is most likely that the turbulence generated outside due to the ground heated by the sun was carried though the dome and across the telescope beam. Finally, when we closed the East vents the $C_n^2$ decreased to $10^{-15}$~$10^{-14}m^{-2/3}$ with relatively strong variations. This simple experiment shows that opening dome vents indiscriminately can be detrimental to dome seeing, and as a first suggestion, it would be advisable to open downwind vents only. However, this obviously needs further testing and validation at nigh time.

We then moved the sensor to a pole on the mezzanine to be off the telescope for nighttime tests, and had it running autonomously for 33h from the evening of October 9th, 2018 until the morning of the 11th, See Figure 9. Although we clearly see different levels of turbulence in the dome between day and nighttime, and there are also trends during the night, the sensor was too far from the telescope beam (and in fact probably too close to the wall of the dome) to be representative of the turbulence experienced by the telescope beam. Indeed, we tried comparing the image quality recorded by the SITELLE FTS spectro-imager[10] at the Cassegrain focus of the telescope on the night of October 9th with the MKAM DIMM and AIRFLOW measurements, as one might expect that IQ-DIMM = AIRFLOW. We do see that at the very end of the night, the SITELLE IQ blows up and so do the AIRFLOW measurements, but this could be due to dome movements.

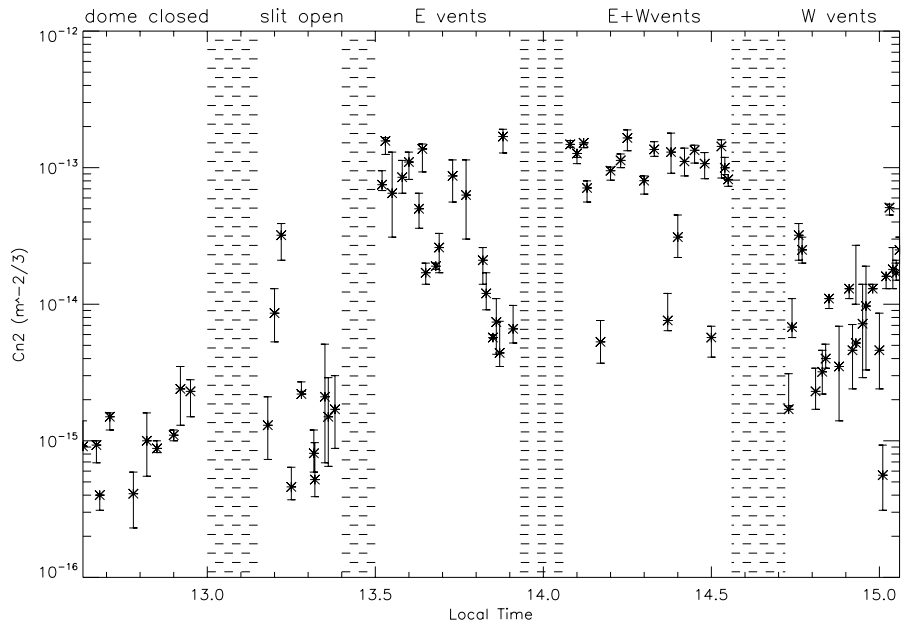

Figure 8. Measurements of $C_n^2$ during a dome vents opening and closing sequence (see text for details). In this and all subsequent plots, the vertical bars show the min and max value of $C_n^2$ along the major and minor axis of the Phase structure function, not measurement errors. Small bars mean isotropic turbulence.

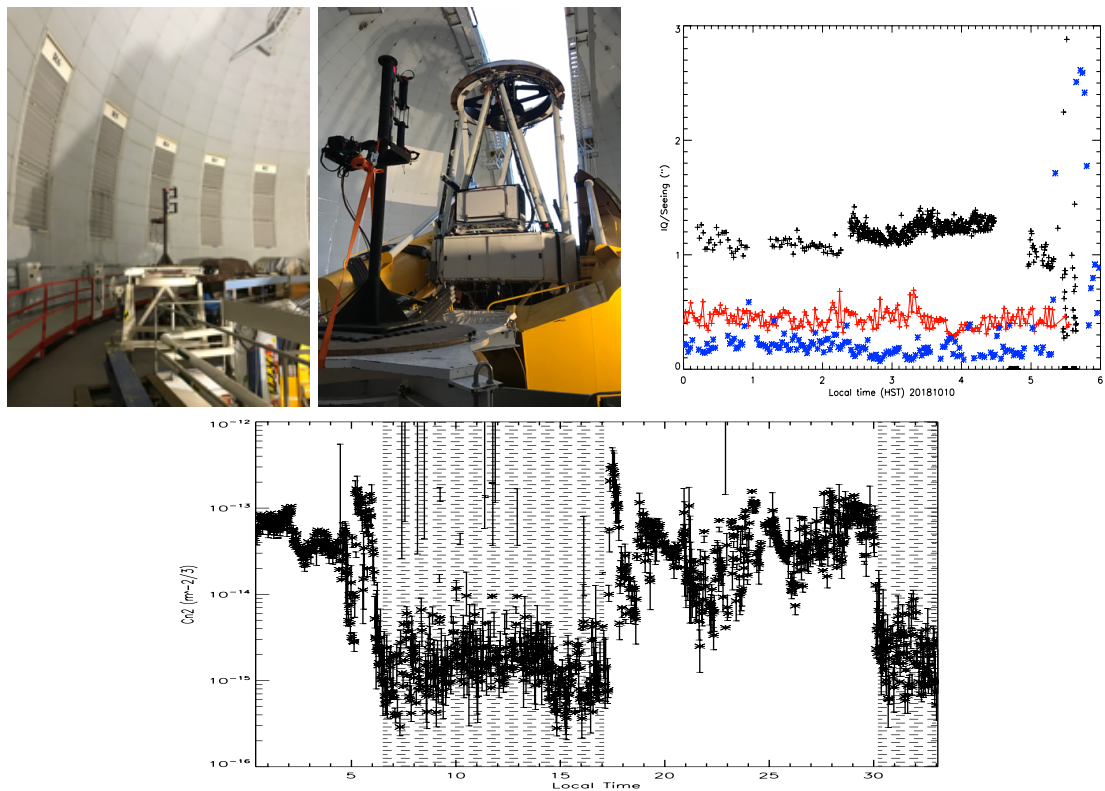

Figure 9. Sensor on CFHT's mezzanine (top right and middle) providing 33 hours of autonomous operation (bottom) showing a clear difference between night and day (hatched) conditions. Comparison of telescope IQ (black), DIMM (red) and AIRFLOW (blue) in the top right panel is inconclusive due to AIRFLOW being so far from telescope beam.

## 3.2 UH88 Daytime testing

We then moved the sensor to the UH88" telescope for further daytime testing, specifically to look for turbulence in the telescope tube. We lowered the sensor into the tube with a rope on two occasions and found consistent results, shown on Figure 10.

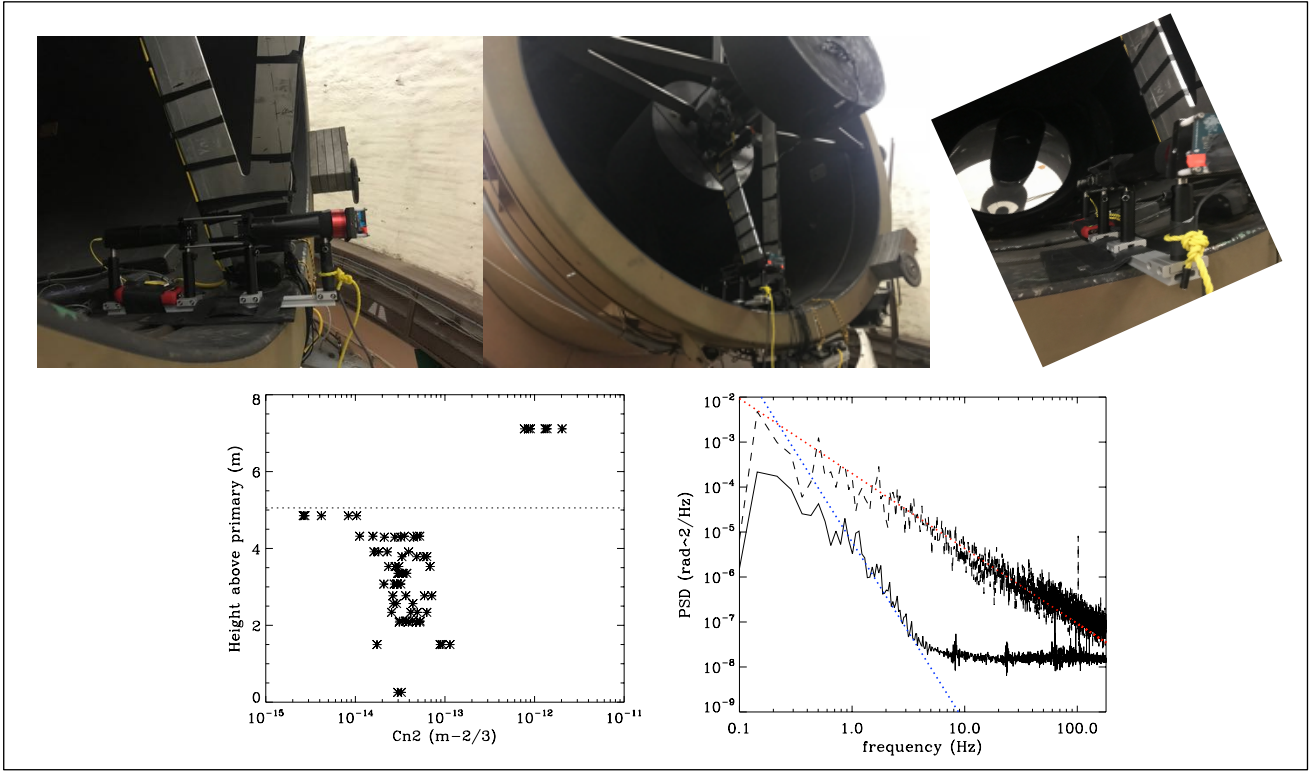

Figure 10: Daytime testing in UH88 tube. On the bottom right plot, we can see stratification of the $C_n^2$ inside the tube, with a gradient towards lower turbulence close to the top of the tube, where there is presumably more mixing of air at different temperature; the primary mirror is at 0m and the top of the tube is at 5m. The points at 7m were at the slit, which was open for these tests. On the bottom right, we show a comparison of the temporal power spectrum between measurements in the slit, fitted with the dotted red line, which is proportional to $f^{-5/3}$ and close to the primary, blue dotted line, proportional to $f^{-4}$.

The $C_n^2$ we found inside the tube was strongest close (within a few meters) to the primary mirror and decreased closer to the top of the tube where there was presumably more mixing between air inside the tube and in the dome. However, the temporal characteristics of these two regimes were very different, with a temporal power spectrum $\propto f^{-4}$ inside the tube. This is what one would expect if the movement of the air was not Kolmogorov but pure tip-tilt, which could be consistent with stable layered air masses at different temperatures, sloshing around the tube.

## 3.3 UH88 imaka tests

For our imaka observing run of February 19th-25th, 2019, we installed two AIRFLOW sensors, one close to the primary mirror, the other on the secondary mirror spider, to try to correlate such measurements with the characteristic echoes of tube seeing. We also installed a camera looking at extra-focal images to try to detect boiling or flying shadows. Unfortunately, we were plagued by foul weather for all the nights but the last and the MKAM MASS-DIMM was not working that night. We found that both sensors followed each other closely with an offset, the sensor close to the open end of the tube having greater values of $C_n^2$, opposite to our daytime tests, We had to close the dome on a few occasions during the night, and each time we observe an asymptotic decrease of the turbulence with the dome over the course of a few hours. While the dome was open, the $C_n^2$ was approximately $10^{-13}m^{-2/3}$ at the primary mirror and $10^{-12}m^{-2/3}$ at the top of the tube. At one point during the night, two of us had to go up to the Cassegrain environment to change a filter, and our AIRFLOW sensors were able to detect the turbulence induced by the warm bodies. Unfortunately, we were unable to detect any correlation between the AIRFLOW sensors and the turbulence profiles provided the imaka wavefront sensors during the intermittent hour and a half when imaka was observing that night. It is possible that the orographic ground layer turbulence was strong that night, masking the dome and tube seeing.

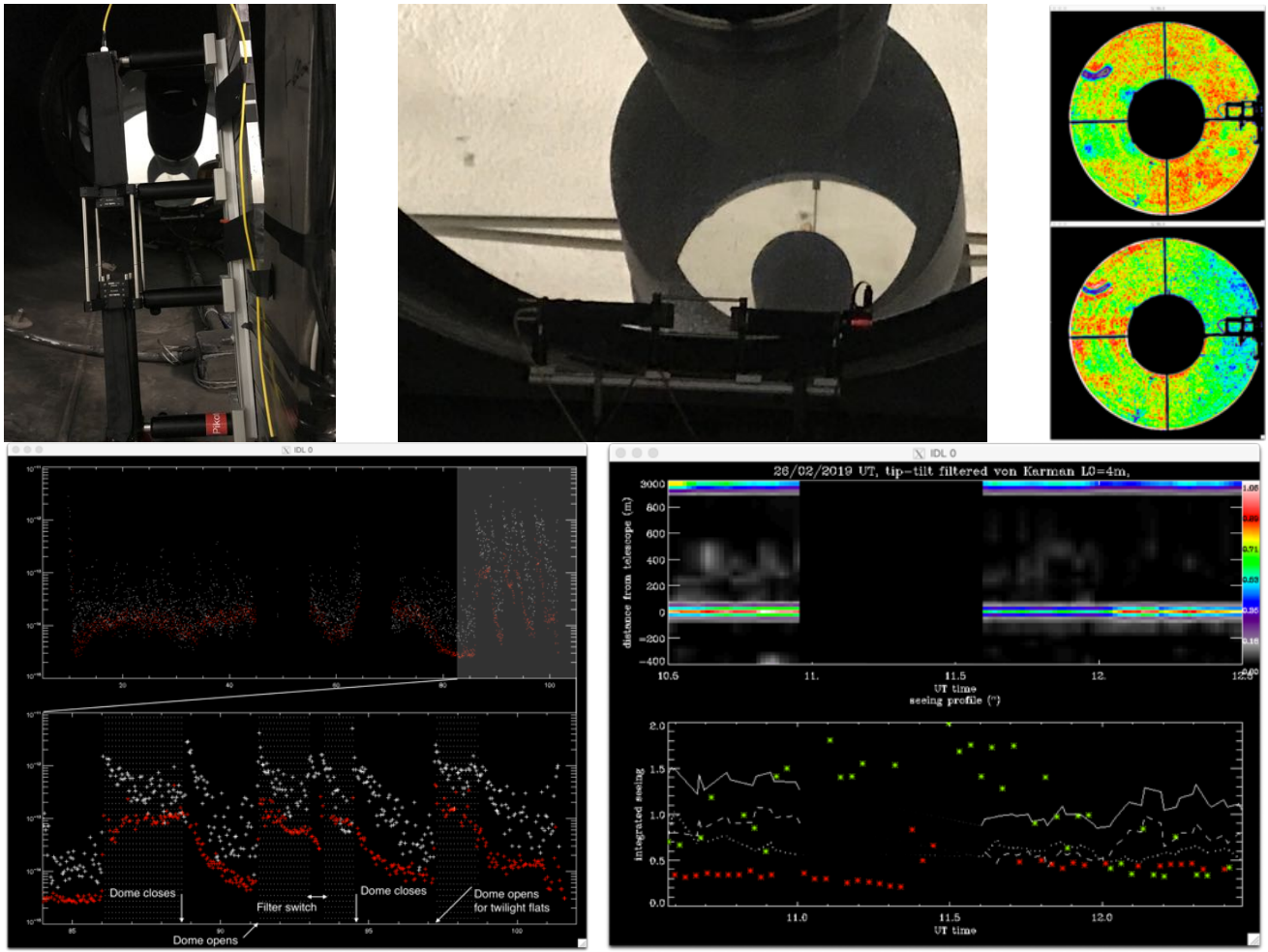

Figure 11: AIRFLOW sensors on the telescope spider and above the primary mirror, the shadows of which can be seen on the extra-focal images. The top extra-focal image show the variance of intensity, while the bottom one is the average intensity; notice the strange plume visible in the North-West quadrant visible on both. Bottom left, the top sensor (in white) is consistently higher than the bottom (in red).The top plot shows the 5 nights of data (with a few gaps when we lost communication with the sensors) and we zoomed in on the last night on the bottom plot. Moments when the dome is open and closed are indicated and can be seen by a sharp rise and asymptotic decay respectively. Bottom right shows the turbulence profile measured by imaka; there doesn't seem to be an echo signature throughout the observing. At the bottom, the full line is the integrated seeing seen by imaka, the dashed line is the ground layer and the dotted line is the free atmosphere (notice the gap in observing when the filter was switched on the science camera). The red and green stars are the bottom and top sensor respectively, which show little correletaion with the ground layer seeing (though they both do increase when people were present on the platform).

### 3.4 Comparison with INTENSE

We had been wanting to compare our sensor with other optical turbulence sensors, such as the INTENSE[11] (Indoor TurbulENce Sensor) which has been well characterized. In May 2019, INTENSE was in the dome of the lunar laser ranging telescope MéO at Plateau de Calern, so we carried out a series of comparative tests. Unfortunately, it was raining the day we had scheduled access to the dome, so we couldn't open it to change the turbulence environment. Instead we used a heat gun to generate very localized turbulence. INTENSE propagates 4 beams along a 2.5m sampling cell, which is ~2m tall;

it measures the relative displacements of the spots on a camera to extract the longitudinal and transverse $r_0$ for different baselines. AIRFLOW sampling cell is much smaller, but when the dome was quiet (i.e. heat gun off), both instruments agreed on the absolute values of $C_n^2$ (on the order of $10^{-14} m^{-2/3}$), providing confirmation that both sensors measure the same thing and are well calibrated. However, when the heat gun was on, our results do show some local effects, related to the different temporal and spatial scales being measured: we started off with AIRFLOW attached to the top beam of INTENSE but because the heat gun was on the floor, the turbulent mixing took place mostly on the lower beams of INTENSE and AIRFLOW did not catch all of it. In the second part of the day, we put AIRFLOW on the bottom beam of INTENSE (as shown in Figure 12, left) and this time, AIRFLOW measured stronger turbulence than INTENSE, but this is understandable, as INTENSE provides an average measurement of the turbulence throughout its cell, so the heat gun's turbulence was diluted.

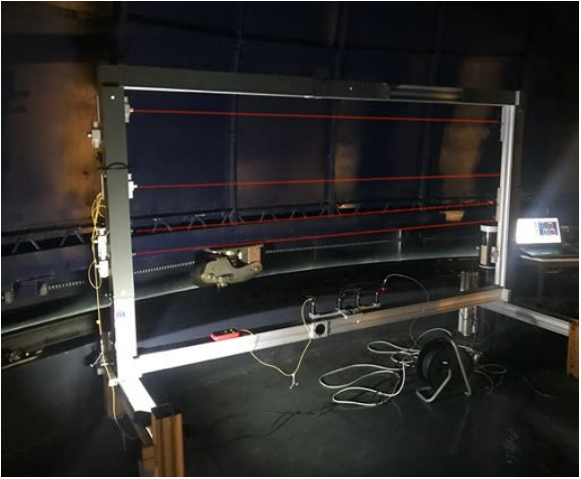 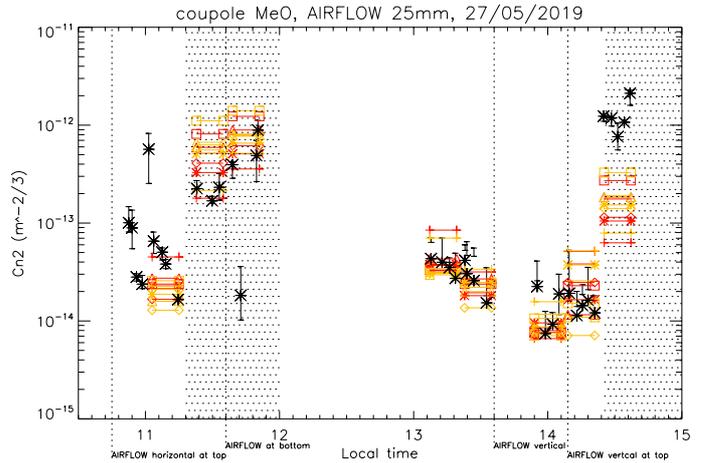

Figure 12: Left, the INTENSE instrument, with AIRFLOW on its lower beam. Right, comparison of AIRFLOW (black crosses) with INTENSE longitudinal (red) and transverse (orange) for its different baselines (various symbols). The INTENSE measurement sequence lasts 10 minutes, which is shown as the horizontal extent of data. Times when the heat gun was on are shown as hatched. The slight discrepancy at the very beginning was due to our own movements inside the dome showing how sensitive AIRFLOW is.

### 3.5 SPEED High Dynamic Range testbench

Because AIRFLOW is so compact, it can also be used to measure the turbulence on laboratory optical benches, so we carried out an experiment on the SPEED[12] (Segmented Pupil Experiment for Exoplanet Detection) high dynamic range testbench. Cameras inside the bench have fans or use liquid nitrogen for their cooling and it is important to know what part of the stability of the high contrast experiment is intrinsic to segment diffraction and stability, and which part is environmental. We installed AIRFLOW inside the enclosure on May 23rd, 2019 and carried out various tests, turning the camera's fan on, etc. However, AIRFLOW was designed to measure turbulence, not the absence thereof, and we were reaching the sensitivity limit of the instrument. Because the turbulence inside the testbench was slow, we were able to improve our sensitivity somewhat by averaging 40 frames at a time. However, the next day, we replaced AIRFLOW with a version with a 50mm diameter beam. This version had previously been deemed less useful for field experiment due to its flexures and vibrations, but the SPEED testbench environment introduced neither and so with a longer sampling cell (500mm), we were able to reach sensitivities down to $10^{-17} m^{-2/3}$. We are currently trying to correlate these turbulence measurements with temperature data from sensors inside the bench. For the longer term, we are developing a specific device, designed to be as highly sensitive to turbulence as possible, to be installed permanently in the testbench.

## 4. LOCAL SEEING

Local seeing is important to every telescope. Local topography can introduce turbulence at ridges or breaks in the terrain, especially if some parts are differentially heated during the day or cool radiatively at night[13]. Hydrodynamic simulations can provide much insight into the mechanical turbulence, but again, lacking reliable thermal input, it is not obvious how to translate such mechanical into optical turbulence. Telescope domes do radiatively supercool and interact with the airflow; this can generate wake turbulence, which could presumably affect the seeing when pointing downwind.

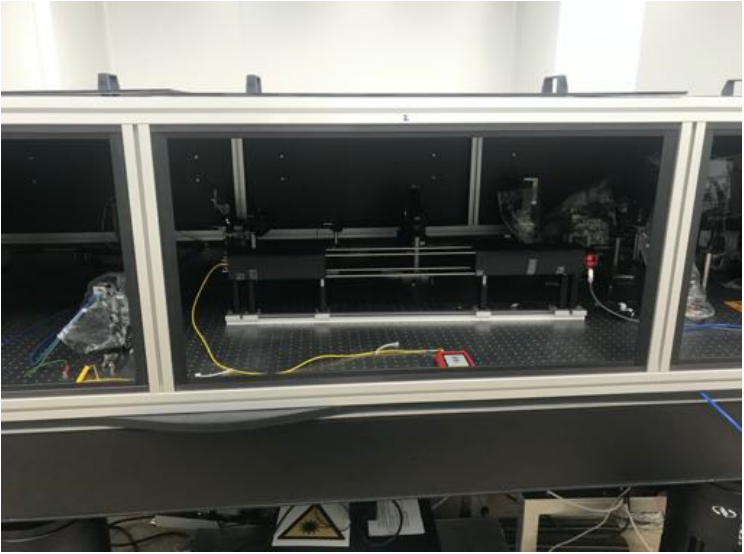
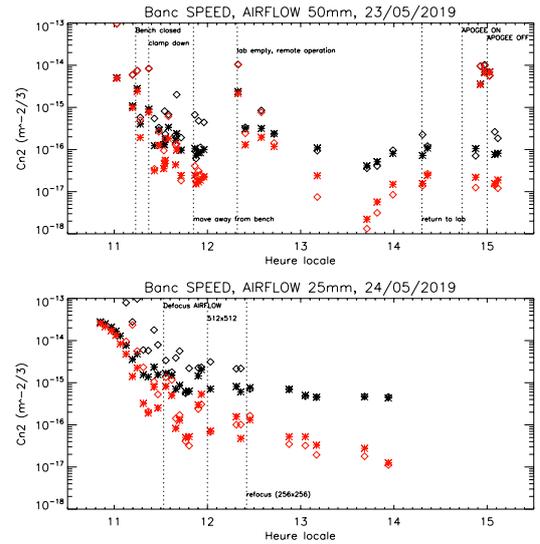

Figure 13: 50mm diameter beam AIRFLOW (with 500mm sampling cell) inside the SPEED testbench. On the top right, results with the 25mm AIRFLOW with camera fans turned on and off. Crosses are full data, diamonds are for tip-tilt filtered, Black is for full frame rate (200Hz), red is for 40 frame averaging (effectively 5Hz).

We devised the mWFS experiment to study such effects: it consisted of 5 high order (30x30) wavefront sensors at CFHT's prime focus observing a 0.5º field in the Pleiades, providing turbulence profiles with vertical resolution on the order of 10m, as we wanted to attempt to separate turbulence inside the dome from the outside ground layer. We expected to see very good ground layer conditions in the first part of the night, looking into the wind (Maunakea's predominant wind is from the North-East), through laminar airflow with no obstacles, and more turbulence after transit, looking downwind into the wake turbulence of the ridge and telescope itself. What we found was the exact opposite, as shown on Figure 14!

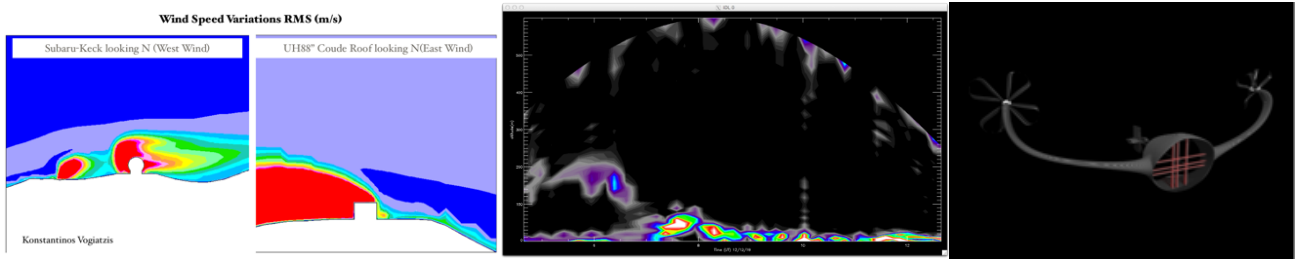

Figure 14: Left, CFD simulation of wake turbulence around buildings from Konstantinos Vogiatzis, showing the effect of ridges. Center: Measured turbulence profile with mWFS on the Pleiades in the course of the night of Dec. 12$^{th}$, 2012. Right: The A in AIRFLOW is for Airborne, and we hope to eventually fly these sensors around building and ridges.

At the start of the night, as we look into the wind, we see a layer 200m above the telescope (so quite high above the ground), that seems to drop as the wind speed drops at around 9pm local time. In the second part of the night, the turbulence is within the very first vertical resolution element, i.e. in the first 10m, with no sign of wake turbulence. A possible explanation for this, suggested by UH meteorologist Steve Businger (private communication) is that the air separates on either side of Maunakea and reconnects in wake turbulence vortices above the summit, and this is what we are looking through. In the second part of the night, the turbulence could be sucked down the leeward descending slope by Bernoulli effect. This indicates that we still have a lot to learn about the local turbulence behavior even outside the buildings. It would be instructive to fly an AIRFLOW (as the first letter of the acronym suggests) using a drone, a balloon or a kite, to scan the volume above the mountain in a variety of conditions!

## 5. CONCLUSIONS AND FUTURE PLANS

AIRFLOW is a reliable and versatile instrument that can quantify optical turbulence inside telescope domes. We are planning on deplying an array of instruments around the CFHT dome by the end of 2019 to perform a site study over the course of a year and provide a prescription to optimally operate the dome vents. We have also carried out preliminary tests at the LBT in Arizona. But more generally, we are trying to gain a better understanding of self-generated turbulence, especially dome and tube seeing, and also even the topographically induced ground layer, with the ultimate goal of correcting it at the source: In the long run, we could consider the use of deflectors, foils, hedges or mesh gratings to improve the mixing (or laminarity) of the air flow and remove most of the ground layer turbulence, even if the idea is not new.

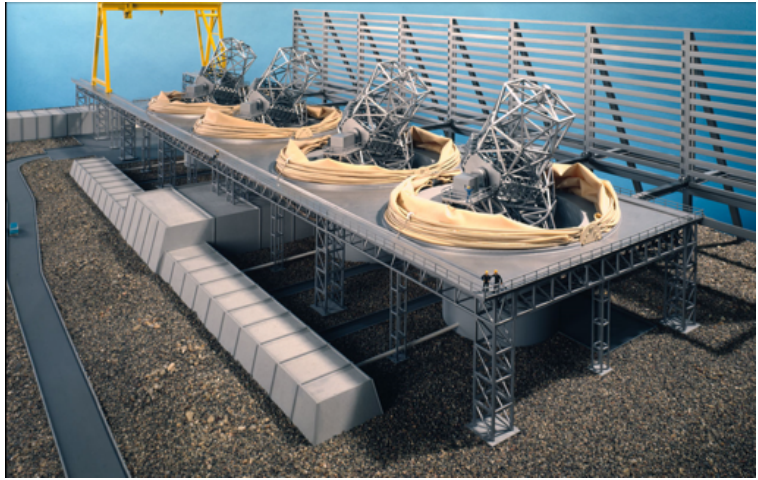

Figure 15: Model of the proposed VLTI from 1984, with fully retractable dome and a gigantic fence in the predominant wind direction to break up the ground layer turbulence.